\RequirePackage{fix-cm}
\documentclass[smallextended]{svjour3}       
\smartqed  
\usepackage{graphicx}
\usepackage{amssymb}
\begin{document}

\title{Sufficient reason and reason enough
}

\titlerunning{Sufficient reason}        

\author{Gustavo E. Romero}


\institute{Instituto Argentino de Radioastronom{\'{i}}a (IAR, CCT La Plata, CONICET) \at
              C.C. No. 5, 1894, Villa Elisa, Buenos Aires, Argentina. \\
              Tel.: +54-221-482-4903 ext. 115\\
              Fax: +54-221-425-4909 ext 117\\
              \email{romero@iar-conicet.gov.ar}
}

\date{Received: date / Accepted: date}

\maketitle

\begin{abstract}
I offer an analysis of the Principle of Sufficient Reason and its relevancy for the scientific endeavour. I submit that the world is not, and cannot be, rational -- only some brained beings  are. The Principle of Sufficient Reason is not a necessary truth nor a physical law. It is just a guiding metanomological hypothesis justified {\it a posteriori} by its success in helping us to unveil the mechanisms that operate in Nature.

\end{abstract}
\vspace{0.3cm}

\begin{quotation}
\begin{flushright}
What necessity would have stirred it up to grow later rather than earlier, beginning from nothing?...Nor will the force of conviction ever permit anything to come to be from what is not.\\[0.5cm]
{\sl Parmenides.}
\end{flushright}
\end{quotation}  

\section{Introduction}
\label{intro}

The Principle of Sufficient Reason (PSR), although famously espoused and advocated by rationalist philosophers such as Spinoza and Leibniz  in the XVII century, has an illustrious history that pervades the whole Western thought (see Schopenhauer 2012). Anaximander is usually credited as the first thinker that invoked the PSR in philosophical reasoning on the occasion of his argument for the earth to be at rest at the center of the universe. Anaximander claimed that since space is isotropic there is no reason for the earth to move in any direction. Then, it does not move. In other words, a symmetry should not be broken if there is not a sufficient reason for breaking it (see McKirahan 1994 on Anaximander's view).  The principle was later adopted by Parmenides, Leucippus and other ancient philosophers. In the Middle Ages, Aquinas argued from the PSR to reject an infinite regress: he held that there must be some reason for the whole chain of causes. Spinoza stated the principle in his famous major work, the {\it Ethics} (Spinoza 1985).  In E1p11d2, we read:

\begin{quotation}
For each thing there must be assigned a cause, or reason, both for its existence and for its nonexistence.
\end{quotation}

\noindent For Spinoza not only there must be a reason for what there is, but also for what there is not. This seems to be a particularly strong version of the principle. 

Leibniz introduced the expression `Principle of Sufficient Reason' and he is its best known exponent and defender. In the {\it Monadology}, sec. 32, he wrote:

\begin{quotation}
There can be no fact real or existing, no statement true, unless there be a sufficient reason, why it should be so and not otherwise, although these reasons usually cannot be known by us. 
\end{quotation}

\noindent And in his second letter to Samuel Clarke, he simplifies (Leibniz and Clarke 2000):

\begin{quotation}
The principle of sufficient reason, namely, that nothing happens without a reason.    
\end{quotation}

\noindent This is not far from the only extant fragment of Leucippus (Taylor 1999):

\begin{quotation}
Nothing happens in vain,  but everything from reason and necessity.    
\end{quotation}

The PSR was under attack in the XVIII century by the empiricists, especially David Hume. Hume critique of causality can be easily extended to sufficient reason. Logical positivists and modern analytic philosophers have also distrusted of the PSR, in part because of its alleged theological implications and in part because its dubious nomological status (see Pruss 2010 for a broad review of contemporary criticisms).  

In this brief paper I want to clarify the role of the PSR in science. I maintain that this principle, properly understood, plays an important role in scientific research. Far from being an obscure tinge from an outdated rationalism in search of theological justifications, I submit that the PSR is a fundamental working hypothesis in the toolkit of any research scientist. But before discussing epistemological issues related to the principle, we should recognise that under the single acronym `PSR' there are a variety of statements of different import and strength, as well as some vagueness that must be dispelled.  Let's see.   

\section{The Principle of Sufficient Reason}
\label{sec:ST}

In his book devoted to the PSR, Alexander Pruss (2010) collects the several ways in which the principle has been enunciated (see also Oppy 2009 for a rational reconstruction). I distinguish four different main forms of the principle (Pruss 2010)\footnote{These forms do not exhaust of course all statements that have been proposed as possible enunciations of the PSR, but are, in my opinion, those more commonly adopted in the philosophical literature.}:
 
\begin{itemize}
\item PSR 1. Everything that is the case must have a reason why it is the case. \\

\item PSR 2. Necessarily every true proposition has an explanation. \\

\item PRS 3. Every event has a cause.  \\

\item PSR 4. {\it Ex nihilo nihil fit} (nothing comes from nothing). \\

\end{itemize}  

These statements are certainly not equivalent. They have presumably different strength and meaning. But in order to compare them and their import, vagueness should be removed from some terms that appear in the statements. Words such as `reason', `explanation', `cause', and `nothing' should be carefully defined in the present context before we can discuss the principle and its place in science. In the next section I proffer elucidations of these terms. 

\section{Reason, explanation, and causality}
\label{sec:2}

The word `reason' is polysemous. I differentiate two main meanings (e.g. Bunge 2003): (1) a mental faculty consisting in thinking in a cogent way, and (2) an ontological justification of the occurrence of an event or a state of affairs. PSR 1 does not refer to properties of the brain, so we better try to refine (2) so as to make of PSR 1 a meaningful statement. An ontological justification for events and states of affairs might be the specification of a sufficient system of causes. In such a case PSR 1 $\rightarrow$ PSR 3. I discuss causes and PSR 3 below. But there is another possible meaning of `ontological justification': the specification of a system of laws and facts such that given a number of conditions A, then the event or state of affairs B follows. For instance, the specification of the law of gravitation and the masses of all objects in the solar system, plus some adequate initial conditions, justify the state of motion of the earth with respect to the sun. In this sense, we can say that there is a `reason' for the earth motion around the sun. I call this type of justification `nomological justification'. Under this interpretation, PSR 1 is not a law, but a metanomological statement (Bunge 1961, 1967). Since laws can be understood as constant relations among properties of things, PSR 1 would be tantamount to 

\begin{itemize}
\item PSR 1a. All events are lawful.        
\end{itemize}

Let us now turn to PSR 2. This version refers to propositions, i.e. classes of statements, and not to the world. As it has been enunciated, it is a statement about our uses of language. Since explanation is an epistemic operation and not a semantic one, in its current form PSR 2 is meaningless. It can be minimally modified, nevertheless, to become a meaningful statement, namely:

\begin{itemize}
\item PSR 2a. Necessarily every true proposition satisfies truth conditions.        
\end{itemize}
  
This is trivially true, but says nothing about the world. Since the proponents of the PSR think that they are saying something about the way the world actually is, we can attempt a different approach replacing propositions by what they represent: facts. We obtain:

\begin{itemize}
\item PSR 2b. Necessarily every fact has an explanation.        
\end{itemize}

I should say a bit more here about explanation if we are going to make any sense of this statement. To explain a fact is to show how it happens. This `to show' means to make explicit a mechanism that produces the fact (see Bunge 2006). A mechanism is a system of processes that occur lawfully. So, PSR 2b can be rendered into:

\begin{itemize}
\item PSR 2c. Every fact results from lawful processes.        
\end{itemize}

Since there is no need for the word `necessary' given the unrestricted universal quantification, I dropped it in this reformulation. PSR 2c is very similar to PSR 1a. If we define facts as either events (changes in the state of things) or states, and admit that events are related to either previous states or other events, then both statements have the same import. 

I turn now to PSR 4. In the formulation given above, PSR 4 is defective since `nothing' is not a thing but a concept: an empty domain of quantification. I propose the following reformulations of PSR 4:

\begin{itemize}
\item PSR 4a. There are not bare facts.        
\end{itemize}

PSR 4a means that all facts are part of a system of facts, the world, where no event occurs isolated. Although this implies a nomic determinism, it is certainly {\em not} a causal determinism, as the one required by PSR 3. Using unrestricted quantification\footnote{$\forall x \;Px \leftrightarrow \neg \exists x \; \neg Px$.} we can rewrite PSR 4a as:

\begin{itemize}
\item PSR 4b. Every fact results lawfully from previous facts.        
\end{itemize}

\noindent This form is quite similar to PSR 2c. 

In order to discuss PSR 3 it is convenient to say a few words about causality before.  

Causation is a mode of event generation (Bunge 1979, Romero and P\'erez 2013). It is not the only way of generating events. Particle decays, as those of elementary particles, such as muons and taus, generate events without causal origination: the existence of no previous event is necessary for the occurrence of these processes. The same is valid for the decay of composed particles such as the various mesons, the neutron, or any spontaneous decay, from radioactive nuclei to the implosion of a star. The events of decay are legal (they occur in conformity to the probabilistic laws of quantum interactions, or other laws or even complexes of laws), but these events and processes are not causal: there is no prior event that we can call the `cause' of these occurrences.  

I suggest the following definition of causal interaction between events\footnote{For more sophisticated definitions see the mentioned works by Bunge (1979) and Romero and P\'erez (2013), as well as Bunge (1977).}: two events $e_1$ and $e_2$ are causally related iff there is at least a process $p$ such that $e_2$ is a component of $p$ and $e_1$ is a component as well, and it is never the case that $e_1$ is not a component of $p$.  Then I say that $e_1$ is a cause of $e_2$. The event $e_2$ is an effect of $e_1$. In symbols:

$$
e_1 \rhd e_2.
$$   

The process $p$ involving $e_2$ can never occur without the existence of $e_1$.  The world is legal and determinate, but not strictly causal. There are events that are not causally related and processes that are not causally originated. If these is correct, then PSR 3 is a false statement\footnote{I owe to an anonymous referee the interesting comment, worth to be cited here, that though reasons are not causes, a reason may become a cause in the head of a decision-maker who proceeds to act in accordance with a causal hypothesis – which may explain why Spinoza, Leibniz and others treated both terms as synonymous.}. 

From our analysis of different proposals for the formulation of the PSR we conclude that, once all the terms have been conveniently defined, the different statements collapse into the following one:\\

\begin{itemize}
\item PSR$^*$. Every fact results lawfully from previous facts. \\        
\end{itemize}

In this formulation `fact' means `a state or a change of state of a thing'. An event is a change of state, so a fact is either a state or an event in a thing. I propose PSR$^*$ as the only version of the PSR that is compatible with modern science. 

Before discussing the ontological and epistemological status of PSR$^*$, I will briefly comment on the system of all things, to which the principle is applied. 

\section{The intelligibility of the world}
\label{sec:3}

The PSR is equated sometimes to the statement, likely inspire by Hegel, that  ``reality is rational''. This can adopt occasionally the form (1) ``the world is rational'' or (2) ``the universe is reasonable''. I submit that all these sentences are nonsense.  Reality is a concept: the set of all real entities. As all sets, it lacks of independent existence, it is a fiction, albeit a convenient one. The word `rational',  to the contrary, qualifies a type of behaviour: the one that is guided by reason, i.e. by cogent thinking. Sets do not think, so reality cannot be rational, hence (1) makes no sense.  The world, on the other hand, can be understood as the system of all events; the universe, as the system of all things (see Romero 2013 for a detailed characterisation). Both  world and universe are concrete entities, but the faculty of thinking, and of thinking reasonably and rationally, is not among their known properties. To the best of our current knowledge, only beings endowed with brains of notable plasticity are able to think. It is difficult to understand what would mean for the universe to think, and even more difficult for the world (since the world is changeless -- Romero 2013). At best, sentences (1) and (2) are false statements. In the worst case, they are not even statements. 

Perhaps what is meant by this type of talk about the world is that it is comprehensible for us, humans. This, in turn, means that we can produce conceptual representations of all aspects of the world. Although we can assume we can do that as a guiding methodological principle for our research (``there are not forbidden topics''), there is no certainty, I think, that we will ever be able to develop the conceptual tools for a full representation of the world and the means to test such representations (see Rescher 1999). This should  not be a  hindrance to our attempts at deepening our understanding of reality. It is in this enterprise where the PSR becomes prominent.

\section{The status of the PSR}
\label{sec:4}

I propose that the correct enunciation of the PSR is PSR$^*$:  every fact results lawfully from previous facts. This is a general statement about facts and laws. It is a statement neither necessary nor obviously true. Since it claims that laws cover the whole range of facts, it is a metanomological statement. It is a condition upon law statements: they ought to cover all the realm of reality. The epistemological status of such a statement is methodological: it is guiding principle for generating knowledge. In every situation where apparent brute facts seem to appear, the PSR$^*$ recommends the search for deeper laws. Any working scientist adopts this principle when an apparent inconsistency appears in the data at hand. Instead of simply assuming brute facts, the responsible scientist proposes a revision of the data or, as a last resource, a modification of the accepted ontology. For instance, the non-conservation of energy, momentum, and spin in some particle decays led the physicist Wolfgang Pauli to postulate the existence of the neutrino in 1930. Recently, the apparent violation of special relativity in neutrino experiments led some scientists to speculate about some exotic explanations and, ultimately, to find a mistake in the interpretation of the experimental data due to some systematics not originally taken into account. In these an many other instances of scientific inquiry the researchers are guided by the non-explicit assumption of PSR$^*$: there must be a lawful explanation of each experimental or observational situation. 

Not being the PSR a necessary truth, the theological scruples of some philosophers are groundless. The principle reflects our disposition to solve problems  in science, but cannot be used to make direct predictions. It is too a general statement for that. Predictions can be made from law statements plus a set of conditions obtained from information about particular states of affairs. We cannot infer the existence of something, e.g.  the neutrino, from the PSR alone. The actual process is that we {\em propose} the existence of the neutrino to satisfy a well tested law (e.g. momentum conservation). We are {\em motivated} by the PSR to demand a fully lawful situation. Ultimately, it will be the experiment that will confirm whether the neutrino exists or not. 

It is important to notice that quantum transitions and other intrinsically probabilistic phenomena are not violations of the PSR$^*$ and {\em do not require} any special interpretation of quantum mechanics. Transitions and decays occur in perfect agreement with the law statements of quantum mechanics. Actually, the probabilistic predictions of quantum mechanics are extraordinarily well corroborated, to the point that most of our modern technology is based on them. I cannot think of a worse attempt to rebut the PSR$^*$ than invoking quantum mechanics. Amazingly, some philosophers have tried to do it...in papers written with computers that operate in accordance with the laws of quantum mechanics.

\section{Closing remarks}
\label{sec:5}

The Principle of Sufficient Reason is a metanomological statement that provides a useful guide in the pursuit of scientific knowledge. It is not a law of nature nor a necessary statement. It is, nonetheless, used by scientist in their everyday work, and has been assumed in many of the most important discoveries of science. This is reason enough for a principle of sufficient reason to be well coveted into the toolbox of any serious researcher.    

\begin{acknowledgements}
I thank Mario Bunge and Daniela P\'erez for stimulating discussions.
\end{acknowledgements}


\bibliographystyle{aipproc}   


\newpage

\section*{Gustavo E. Romero} Full Professor of Relativistic Astrophysics at the University of La Plata and Chief Researcher of the National Research Council of Argentina. A former President of the Argentine Astronomical Society, he has published more than 300 papers on astrophysics, gravitation and the foundations of physics. Dr. Romero has authored or edited 9 books (including {\sl Introduction to Black Hole Astrophysics}, with G.S. Vila, Springer, 2013). His main current interest is on black hole physics and ontological problems of space-time theories.

\end{document}